\documentclass[smallcondensed]{svjour3}
 \usepackage{amssymb}
 \usepackage{amsfonts,amsmath,epsfig}
 \usepackage{graphicx} 
 \usepackage{palatino}
\usepackage{natbib}
\usepackage{epstopdf}
\usepackage{siunitx}
\usepackage[percent]{overpic}
\def\rr#1{(\ref{#1})}
\def\sech{\textrm{sech}}
\newcommand{\be}{\begin{equation}}
\newcommand{\en}{\end{equation}}

\newcommand{\flex}{\mathfrak{F}}

\def\pafrac#1#2{\frac{\partial #1}{\partial #2}}
\def\p2frac#1#2{\frac{\partial^2 #1}{\partial #2^2}}

\def\kr{\tilde{\kappa}}
\def\tbk{\tanh(\beta \kappa)}
\def\tbk0{\tanh(\beta \kappa_0)}
\def\tbz{\tanh(b z)}

\DeclareSIUnit{\molar}{M}

\DeclareSIPostPower\tothefourth{4}

\begin{document}
\title{Curvature-sensitive kinesin binding can explain microtubule ring formation and reveals chaotic dynamics in a mathematical model}
\author{S. P. Pearce  \and M. Heil \and O. E. Jensen \and G. W. Jones  \and A. Prokop}

\institute{S. P. Pearce \and M. Heil \and O. E. Jensen \and G. W. Jones 
\at School of Mathematics, University of Manchester, UK
\and
S. P. Pearce \and A. Prokop
\at Faculty of Biology, Medicine and Health, University of Manchester, UK
}

\titlerunning{Curvature-sensitive kinesin binding can explain microtubule ring formation and reveals chaotic dynamics}

\maketitle
\begin{abstract}
Microtubules are filamentous tubular protein polymers which are essential for a range of cellular behaviour, and are generally straight over micron length scales. However, in some gliding assays, where microtubules move over a carpet of molecular motors, individual microtubules can also form tight arcs or rings, even in the absence of crosslinking proteins. Understanding this phenomenon may provide important explanations for similar highly curved microtubules which can be found in nerve cells undergoing neurodegeneration. We propose a model for gliding assays where the kinesins moving the microtubules over the surface induce ring formation through differential binding, substantiated by recent findings that a mutant version of the motor protein kinesin applied in solution is able to lock-in microtubule curvature. For certain parameter regimes, our model predicts that both straight and curved microtubules can exist simultaneously as stable steady-states, as has been seen experimentally. Additionally, unsteady solutions are found, where a wave of differential binding propagates down the microtubule as it glides across the surface, which can lead to chaotic motion. Whilst this model explains two-dimensional microtubule behaviour in an experimental gliding assay, it has the potential to be adapted to explain pathological curling in nerve cells. 
\end{abstract}

\section{Introduction}
The skeleton of cells (cytoskeleton) is essential for cell structure, dynamics and function. It is formed by long filamentous protein polymers of three different classes: actin, intermediate filaments and microtubules (MTs). Of these, MTs are the stiffest filaments with important roles in cellular processes, such as cell motility, division, organisation, adhesion, signalling and intracellular transport. MTs are composed of $\alpha$- and $\beta$- tubulin heterodimers which are bonded in a polar head-to-tail fashion to form long chains known as protofilaments; these protofilaments are then assembled into a helical tube. For a detailed description of how microtubules behave, see for example \citet{hawkins2010} or \citet{barsegov2017}.

Many different proteins bind to MTs, controlling MT behaviours, including their nucleation, (de)- polymerisation, stabilisation, severing, biochemical modification, and crosslinking to each other or other cellular components \citep{lawson2013, prokop2013}. One particular class are MT-associated motor proteins, which use ATP as an energy source to walk along MTs, either to slide them against each other or to use MTs as intracellular highways to transport cargo around cells. Two fundamentally different classes of MT-associated motor proteins exist: the various members of the kinesin family of which most walk towards one end of the MT, and the dynein/dynactin complex which moves towards the other  \citep{prokop2013, schliwa2003}. 

\begin{figure}[thbp]
\includegraphics[width=0.8\textwidth,keepaspectratio]{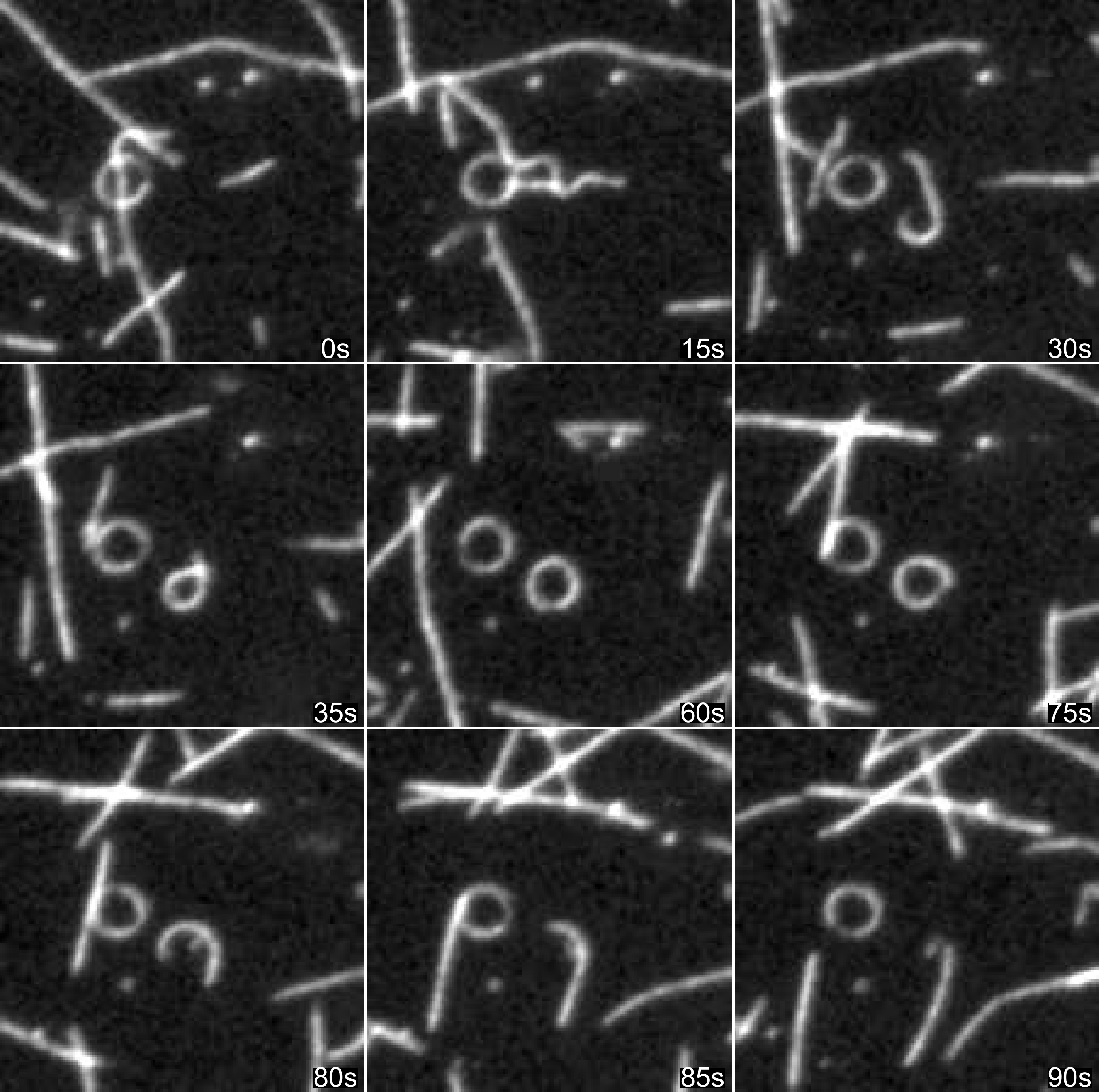} 
\caption{Time-series of an initially straight MT forming a loop at \SI{35}{\second}, rotating until \SI{75}{\second} before re-straightening, extracted from Supplementary Movie 1 of \citet{liu2011}. Another adjacent MT stays in a loop through the whole video. Each frame is \SI{15}{\micro \meter} square. Used with permission of Prof. J. Ross.}
\label{ringpics}
\end{figure}

Outside of cells, a powerful \textit{in vitro} tool to study MT behaviour is a gliding (or motility) assay. In these experiments, motor proteins (typically kinesin-1) are adsorbed onto a solid surface in a drop of solution. When MTs are added, the surface-attached motor proteins attempt to walk along them, causing the MTs to glide over the surface. Typically in these assays, MTs stay relatively straight, as would be expected from their large persistence length (\SIrange{2}{4}{\milli\metre} \citep{howard2001}). However, in certain experimental conditions, MTs can form micron-sized rings; such conditions include high MT density or the presence of an air-medium interface \citep{weiss1991, amos1991, liu2011, kawamura2008, kabir2012}. Strikingly, these MTs are able to transform from straight gliding to a curved circling motion and back again \citep{liu2011}, showing a dynamic and reversible ability to change curvature, implying that this is not due to permanent damage or irreversible damage/repair cycles \citep{schaedel2015} (see Figure \ref{ringpics}). 

Studying the mechanisms that underlie MT curling has important applications. For example, systems based on MT-kinesin gliding assays have potential uses as lab-on-a-chip medical devices, utilizing the ability to bind only selected proteins to MTs through the choice of specific cargo adapters, leading to advective transport rather than mere diffusion \citep{bachand2014, chaudhuri2017}. These nano-devices need to be robust for potential clinical uses, but the presence of MT rings may disrupt their design. 

Furthermore, curved MTs as observed in gliding assays are similarly found in cells, particularly in axons. Axons are the cable-like extensions of nerve cells; their structural backbone is formed by straight, parallel bundles of MTs. However, in the ageing brain or in nerves affected by certain neurodegenerative diseases (e.g. some forms of motor neuron disease), MTs are found to curl up with similar diameters as observed in gliding assays \citep{sanchez2009, voelzmann2016, voelzmann2017}. 

To explain this phenomenon, the model of local axon homeostasis has been put forward \citep{voelzmann2016}. It proposes that MTs in axonal environments have a strong tendency to curl up – likely due to high abundance of MTs and MT-associated motor proteins, thus meeting the conditions known to cause rings in gliding assays. Various MT-regulating proteins are required to \lq tame' MTs into ordered bundles; functional loss of these regulators increases the risk of MT curling and could explain neurodegeneration linked to them \citep{voelzmann2016}. This model represents a paradigm shift for the explanation of certain forms of axon degeneration, by putting the emphasis on MTs as the key drivers of axon decay.  

To lend credibility to this model, it is pivotal to identify and validate the mechanisms that can explain the phenomenon of MT curling. So far, \citet{ziebert2015} introduced a model to explain the formation of MT rings which suggests that, in the presence of the MT-stabilising drug taxol, each tubulin dimer may exist in two distinct conformations, one slightly shorter than the other. In their model, protofilaments are able to switch between these two states; when only some of the protofilaments are switched this leads to a longitudinally curved MT as an energetically favoured condition, providing a mechanism to create rings via an internal change to the MT. 

\begin{figure}[thbp]
\includegraphics[width=0.5\textwidth,keepaspectratio]{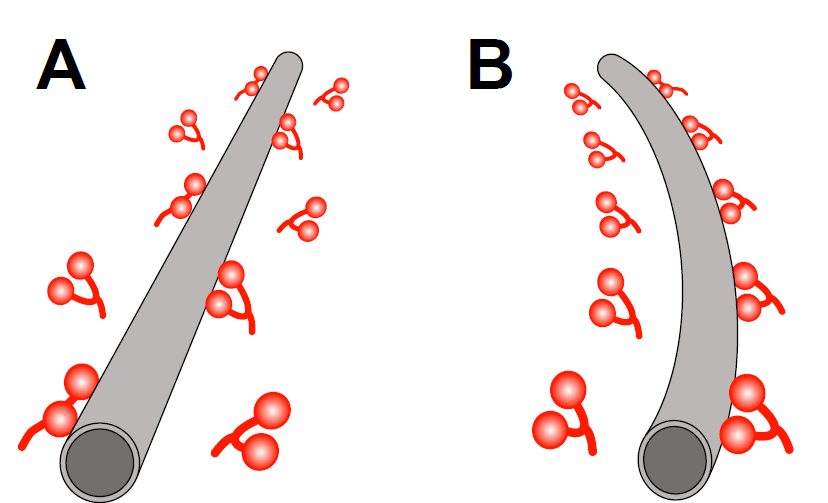} 
\caption{Sketch of the proposed model. A) When the MT is straight, the surface-bound kinesin is equally likely to bind to each side, with no effect on the overall curvature. B) If a MT becomes curved, the likelihood of kinesin binding from each side becomes asymmetric; this asymmetry in the amount of bound kinesin to each side induces curvature by acting as a lateral reinforcement. As the MT then continues to glide across the array, the newly encountered kinesin will also preferentially bind to the same side.}
\label{sketch}
\end{figure}

Here, we explore the complementary possibility that differential binding of external factors can actively contribute to MT curling. \citet{peet2017} show that MTs which are being bent in a flow chamber normally straighten after the flow is removed, but stay curved in the presence of a non-motile version of kinesin-1. They propose that this non-motile kinesin has a tendency to bind preferentially to the convex side of curved MTs and, by doing so, stabilise them in bent confirmation (see Figure \ref{sketch}); at higher concentrations this behaviour disappears, presumably because oversaturation occurs so the kinesin binds in equal amounts on all sides of the MT. 

This behaviour is consistent with findings for other MT-associated proteins, in particular tau \citep{samsonov2004} and doublecortin \citep{bechstedt2014, ettinger2016}, which bind differentially between straight and curved MTs due to conformational changes that happen on the structural scale of the individual tubulin dimer: at a curvature of $\SI{1}{\per \micro \meter}$ the tubulin dimer spacing at the outside of the MT is 2.5\% larger compared to that of the inside. 

Here we present a model based on the hypothesis that curvature-selective binding can occur in MT-kinesin gliding assays; the flexible neck linker of the surface-attached kinesins can extend up to \SI{45}{\nano \metre} from the surface \citep{palacci2016}, and is therefore long enough to reach the curved sides of the MT which are typically held at around \SI{17}{\nano \metre} above the surface \citep{kerssemakers2006}). 
Our model reproduces key behaviours of MTs observed in gliding assays, with a bistable regime where straight MTs and MT rings can coexist, and predicts how they can be controlled. Furthermore, we find unsteady propagating wave solutions and chaotic dynamics within the system, which have not been previously reported for filaments and may reflect true MT behaviours that have escaped the attention of experimenters so far.

\section{Mathematical Model}
\subsection{Filament Dynamics}
We consider a MT as an inextensible filament represented by a curve, $\mathbf{x}(s,t)$, parametrized by its arclength $s$ measured at time $t$, which lies within a two-dimensional plane. This is a reasonable approximation for gliding assays, where the MTs remain close to the surface throughout their movement. The treatment shown here follows the standard case where the filament has no reference curvature \citep[see, for example,][]{audoly2015, ziebert2015, canio2017}. 

As the individual tubulin dimers are of fixed length, the total arclength is constant and we impose the inextensibility constraint, $\mathbf{x}' \cdot \mathbf{x}'=1$, where a prime denotes differentiation with respect to $s$. Relative to fixed Cartesian coordinate axes $\mathbf{e}_x, \mathbf{e}_y$, we define the tangent vector as
\begin{equation}\label{tequation}
\mathbf{t} = \mathbf{x}' = \cos \theta \mathbf{e}_x + \sin \theta \mathbf{e}_y,
\end{equation} 
where $\theta(s)$ measures the angle between $\mathbf{t}$ and $\mathbf{e}_x$. The normal vector $\mathbf{n}$ is then given by the relation $\mathbf{t}' =\kappa \mathbf{n}$, which defines the curvature $\kappa = \theta'$.

We will assume that the filament has a variable reference curvature (to be specified later), $\kr(s,t)$, and that the mechanical energy of the MT is a function of the squared deviation of the curvature from the reference curvature,
\begin{equation}\label{energy1}
E = \frac{1}{2}\int_0^L \left[ B (\kappa-\kr)^2 + \lambda \left(\mathbf{x}' \cdot \mathbf{x}' -1\right) \right]\;ds.
\end{equation}
Here $\lambda$ is a Lagrange multiplier enforcing the inextensibility constraint and $B$ is the bending (flexural) modulus. Taking the variational derivative of \rr{energy1}, utilizing $\delta \kappa = \mathbf{n} \cdot \delta \mathbf{x}''$, we find 
\begin{align}\label{varderiv}
\delta E &= \int_0^L B (\kappa - \kr) \delta \kappa + \lambda (\mathbf{x}'\cdot \delta \mathbf{x}')  \; ds \nonumber\\
&= \left[B (\kappa-\kr) \mathbf{n} \cdot \delta \mathbf{x}'\right]_0^L + [(\lambda \mathbf{x}' - B ((\kappa-\kr) \mathbf{n})') \cdot \delta \mathbf{x}]_0^L \nonumber\\ 
&+ \int_0^L  ((B (\kappa-\kr) \mathbf{n})' - (\lambda \mathbf{x}') )' \cdot \delta \mathbf{x} \; ds
\end{align}
where $\delta$ represents the variation of a quantity and $\mathbf{n}' = -\kappa \mathbf{t}$. The elastic force density, $\mathbf{f}$, acting on an element of the MT is therefore given by 
\begin{equation}
\mathbf{f} = -\frac{\delta E}{\delta \mathbf{x}} = - B ((\kappa-\kr) \mathbf{n})'' + (\lambda \mathbf{x}')',
\end{equation}
subject to $\mathbf{x}' \cdot \mathbf{x}'  = 1$. Here $\mathbf{f}$ has units of force per unit length, and may be considered as the circumferentially averaged surface stress \citep{lindner2015}.
The MT is immersed within a viscous medium at very low Reynolds number, and so we use resistive force theory, a Stokes-flow approximation which takes advantage of the aspect ratio being small, $\epsilon = h/L \ll 1$, where $h$ is the MT diameter (\SI{25}{\nano\meter}). This is the simplest approximation of slender-body theory, and gives the local dynamic relation \citep{lindner2015}, 
\begin{equation}\label{forcevel}
\mathbf{v} = \frac{c}{2 \pi \mu} \mathbf{P} \cdot (\mathbf{f}+\mathbf{f}_{ext})
\end{equation}
where $\mathbf{v} = \pafrac{\mathbf{x}}{t} = \dot{\mathbf{x}}$ is the velocity of material points, $\mathbf{f}_{ext}$ is the external force per unit length acting along the MT, $\mu$ is the fluid viscosity, and the tensor $\mathbf{P} \equiv (\mathbf{I} + (\xi-1)\mathbf{t}\mathbf{t}) = \mathbf{n}\mathbf{n} + \xi \mathbf{t}\mathbf{t}$ reflects the anisotropic drag on the filament due to its shape. The constant $c = \ln (2 \epsilon^{-1})$ is a free-space slender body ratio \citep{becker2001}, and we also use the free-space approximation $\xi=2$ for an idealised slender filament. Both of these neglect the effect of the nearby surface; a more refined approach is likely to predict higher values of overall drag. There is nothing to prevent self-intersection of the filament within this model; if self-intersection does occur the filament is therefore assumed to go out of plane, crossing over or under itself. The equations of motion are therefore given by
\begin{equation}\label{goveqn}
\dot{\mathbf{x}} =\frac{c}{2 \pi \mu} \mathbf{P} \cdot 
\left(- B ((\kappa-\kr) \mathbf{n})'' + (\lambda \mathbf{x}')' + \mathbf{f}_{ext} \right), \quad  \mathbf{x}' \cdot \mathbf{x}' =1, \;\; 0 \leq s \leq L.
\end{equation}
Due to the constraint, this is a ninth-order (in $s$) system of differential algebraic equations with index 3, so six spatial boundary conditions are required to fully specify the system. It is easier to work in terms of intrinsic coordinates which move with the filament, so we take the derivative of \rr{goveqn} with respect to $s$, and use $\dot{\mathbf{x}}' = \dot{\theta} \mathbf{n}$ to give two equations in the normal and tangential directions respectively,
\begin{subequations}\label{intrineqns}
\begin{align}
\dot{\theta} &= \frac{c}{2 \pi \mu} \left( -(B K'' - \tau \kappa)' + \xi (\tau' + B \kappa K'  + f_m)\kappa \right)\\
0 &= \frac{c}{2 \pi \mu}\left( (B K'' - \tau \kappa) \kappa + \xi (\tau' + B \kappa K' + f_m)'  \right) 
\end{align}
\end{subequations}
where $K = \kappa - \kr$ is the excess curvature and $\tau = \lambda + B \kappa K$ is a generalised tension. Here, as in \citet{ziebert2015}, we have assumed that the external force comes solely from the action of the kinesin motors, which force the microtubule along its tangent and so $\mathbf{f}_{ext} = f_m \mathbf{t}$, where $f_m$ is a constant. Equation \rr{goveqn} nevertheless allows the MT to move normal to its centreline. The position $\mathbf{x}$ may then be found from the filament angle $\theta$ by integrating \rr{tequation}.
The natural boundary conditions for a free end of the MT come from the variational principle \rr{varderiv}, and are given by
\begin{equation}
K=0, \quad K'=0, \quad \tau=0.
\end{equation}
For a fixed end we have $\mathbf{x}=\mathbf{x}_0$, and therefore we set $\dot{\mathbf{x}} = 0$ in \rr{goveqn} which leads to the force conditions,
\begin{equation}
\tau \kappa - B K'' =0, \quad \tau' + B \kappa K' +f_m =0,
\end{equation}
which are supplemented with either $K=0$ for a freely rotating pinned end or $\theta =\theta_0$ for a clamped end.

For a straight filament with $\kappa = \kr = 0$, \rr{goveqn} gives
\begin{equation}\label{glidevel}
v=|\dot{\mathbf{x}}| = \xi \frac{c f_m}{2 \pi \mu},
\end{equation}  
which allows us to estimate an appropriate tangential force being applied from measurements of the MT velocity. Typical velocities for gliding assays are \SIrange{0.5}{1}{\micro \metre \per \second}, with differences due to factors such as viscosity, ATP concentration, temperature and salt concentrations. While it would be reasonable to expect that the speed of the MT would be proportional to the amount of kinesin attached to the MT (and hence the surface kinesin density), as equation \rr{glidevel} implies, this is not seen in experiments, which show that the velocity is constant for kinesin densities of \SIrange{10}{10000}{\per \micro \metre \squared} \citep{howard1989}. We therefore will use $v$ to set an appropriate choice of $f_m$.

\subsection{Kinesin Binding}
We now turn to the binding of the surface-bound kinesin to the MT, which we will model as a continuous field, assuming that the concentration is sufficiently high for this to be valid. Here we focus on the \lq sides' of the filament, arbitrarily denoting them with $+$ and $-$ with associated bound concentrations $c^+$ and $c^-$; we do not model the protofilaments which are directly above the surface as we assume that they will not affect the reference curvature. The proteins bind and unbind to the filament according to standard protein binding kinetics, 
\begin{equation}\label{concgoveqn}
\dot{c}^\pm + v_s c^{\pm \prime} = a^\pm (\kappa) \; \left(1 - \frac{c^\pm}{c_{max}}\right) - \delta c^\pm + D c^{\pm \prime \prime},
\end{equation}
where $a^\pm (\kappa)$ is a curvature-dependent association rate, $\delta$ is a disassociation rate, $c_{max}$ is the maximum number of binding sites per unit length and $D$ is a diffusion constant (measured as \SI{0.036}{\micro\metre\squared\per\second} for kinesin-1 \citep{lu2009}). The left hand side of \rr{concgoveqn} is a material derivative, incorporating the fact that we are working in intrinsic coordinates while the kinesin is fixed to the surface, where $v_s$ is the instantaneous MT velocity \rr{goveqn} projected in the tangential direction, 
\begin{equation}\label{vseqn}
v_s \equiv \dot{\mathbf{x}} \cdot \mathbf{t} = \xi (\tau' + B \kappa K' + f_m).
\end{equation}
Dividing \rr{concgoveqn} by $c_{max}$, we use the bound ratios $\phi^{\pm} = c^{\pm}/c_{max}$ as dependent variables, giving
\begin{equation}\label{concgoveqn2}
\dot{\phi}^\pm= \frac{a^\pm (\kappa)}{c_{max}} \; (1 - \phi^\pm) - \delta \phi^\pm + D \phi^{\pm \prime \prime} - v_s \phi^{\pm \prime}.
\end{equation}
At the ends of the MT, we allow no diffusion-based flux of the protein (although it will \lq fall off' the trailing end with the velocity $v_s$) and hence impose $\pafrac{\phi^{\pm}}{s} = 0$ at both ends. 
Although we are modelling the MT as a one-dimensional rod, in reality it has a complex protein structure, with each tubulin monomer consisting of approximately 450 amino acids folded into a 3D arrangement with charged residues protruding from the surface. Bending the entire filament moves these residues in relation to each other, expanding those on one side and contracting those on the other; such changes can be expected to change the binding kinetics of associated proteins, as these are also complex charged structures. 
The precise nature of this relationship is unknown, but here we will assume a sigmoidal relationship of the protein association rate $a$ on the local curvature $\kappa$, 
\begin{equation}
a^\pm (\kappa) = \alpha_0 (1 \pm \tanh(\beta \kappa)),
\end{equation} 
where $\beta$ acts as a scaling factor (with dimensions of length) to determine the degree of preferential binding of the kinesin to curved MTs. This implies that when the MT becomes curved the binding rates to each side of the MT will locally change, and increasing the value of $\beta$ will mean that the differential binding is more sensitive to small curvatures. 

The choice of this sigmoidal relationship ensures that the association rate both saturates at high curvature and that $a^\pm (\kappa)$ is always positive; we have checked that other functional forms can be used to similar effect.

We assume that the average on-rate $\alpha_0$ is proportional to both the number of kinesin molecules available in the vicinity of the MT and their ability to reach one side of the MT, 
\begin{equation}\label{gamdef}
\alpha_0 = d \Gamma \omega_{on},
\end{equation} 
where $\Gamma$ is the surface kinesin density, assumed to be sufficiently large for depletion not to be a concern, $d$ is the maximum distance kinesin can extend (\SI{45}{\nano \metre} \citep{palacci2016}), and $\omega_{on}$ is an attachment rate per kinesin molecule within range, given as \SI{20}{\per \second} \citep{chaudhuri2016}.  We note that this is a high estimate, because we neglect the binding to protofilaments directly above the surface.

Our final model assumption is that the local concentration of bound protein influences the intrinsic curvature of the filament, by acting as a brace on the side of the filament or some other conformational change, as suggested by \citet{peet2017}. As the protofilaments are bonded to each other via lateral bonds, it is assumed that this is able to affect the entire MT. If only one side of the MT has a high concentration this will prevent the filament from straightening, altering the MT reference curvature, while if both sides have bound protein then there will be no net effect on the curvature. Again, we assume a sigmoidal dependence of $\kr$ on the difference between the two \lq sides' of the MT,
\begin{equation}\label{kreqn}
\kr = \kappa_c \tanh \left(\gamma (\phi^+ - \phi^-) \right),
\end{equation}
where $\gamma>0$ is a scaling factor that controls the steepness of the MT response to differential binding and $\kappa_c>0$ is the maximum characteristic curvature. These unknown constants will depend on the precise nature of the bracing effect, but the measured lattice expansion of $1.6\%$ in \citet{peet2017} suggests $\kappa_c=\SI{0.625}{\per \micro \metre}$. The MT will therefore curve towards the side with less bound protein, as shown in Figure \ref{sketch}. 

\subsection{Non-dimensionalisation}
We non-dimensionalise equations (\ref{intrineqns}, \ref{vseqn}, \ref{concgoveqn2}, \ref{kreqn}) with respect to the MT length $L$ and the unbinding time $\delta^{-1}$, resulting in the domain of integration being $s \in (0,1)$, yielding 
\begin{subequations}\label{nondimeqns}
\begin{align}
\chi^{-1} \dot{\theta} &= -(K'' - \tau \kappa)' + \xi (\tau' + \kappa K'  + \flex)\kappa  \\
0 &=  (K'' - \tau \kappa) \kappa + \xi (\tau' + \kappa K' + \flex)'    \\
\dot{\phi}^\pm &= \alpha (1 \pm \tanh(\beta \kappa)) (1 - \phi^\pm) -  \phi^\pm - v_s \phi^{\pm\prime} + D \phi^{\pm\prime\prime},\\
\kr &= \kappa_c \tanh \left(\gamma (\phi^+ - \phi^-) \right), \\
v_s &= \xi (\tau' + \kappa K' + \flex),
\end{align}
\end{subequations}
where the three dimensionless numbers,
\begin{equation}\label{dimensionless}
\flex = \frac{f_m L^3}{B},\quad \chi = \frac{c B}{2 \pi \mu L^4 \delta}, \quad \alpha = \frac{\alpha_0}{c_{max} \delta}
\end{equation}
are the ratio of forcing to bending rigidity, called the flexure number in \citet{iseleholder2015}, the ratio of elastic to viscous forces (inversely related to the Sperm number \citep{lowe2003}) and the updated base on-rate respectively. 
The boundary conditions at a free end are 
\begin{equation}
K= K'= \tau=\phi^{+\prime}=\phi^{-\prime}=0,  
\end{equation}
while at a pinned end we have,
\begin{equation}
\tau \kappa - K'' = \tau' + \kappa K' + \flex = K =\phi^{+\prime}=\phi^{-\prime}=0.
\end{equation}
To connect back to the spatial positions, \rr{goveqn} may be written as,
\begin{subequations}\label{xydot}
\begin{align}
\dot{x} &= \chi (-\sin \theta (K'' +\tau \kappa) + \xi \cos \theta (\tau' + K' \kappa + \flex))  \\
\dot{y} &=  \chi (\phantom{-}\cos \theta (K'' +\tau \kappa) + \xi \sin \theta (\tau' + K' \kappa + \flex) ),
\end{align} 
\end{subequations}
which we can calculate after solving for $\kappa$. We can also use \rr{tequation} to get the shape, supplementing with \rr{xydot} to find the position at a single point.

For the examples shown here, we will set $D=\SI{0.036}{\micro \metre \squared \per \second}$ \citep{lu2009}, $v = \SI{0.5}{\micro \metre \per \second}$, $\kappa_c=\SI{1}{\per \micro \metre}$ (comparable to the \SI{0.625}{\per \micro \metre} suggested in \citet{peet2017};  the exact value does not affect the primary conclusions here), $\delta = \SI{1}{\per \second}$ \citep{chaudhuri2016}. For calculating $\chi$, there is a wide range of measured values of the bending modulus $B$, depending on the measuring technique and conditions, with a noticeable length dependence \citep{pampaloni2006}. Similarly, the viscosity $\mu$ is not clear, as the medium of the gliding assay is more viscous than pure water, and so we shall set $\chi = \chi_0/ L^4$, where $\chi_0 = \SI{1}{\micro \metre \tothefourth}$ or $\SI{3}{\micro \metre \tothefourth}$ in the examples below.

As we are considering only the protofilaments on the sides of the MT, we have $c_{max}= \SI{250}{\per \micro \metre}$ if we only consider two protofilaments as being available for binding on each side. Combined with the estimates of the other values above, this gives $\alpha = 3.6$ when $\Gamma = \SI{1000}{\per \micro \metre \squared}$.

\section{Results}
\subsection{Uniform Steady-States}\label{steadystates}
First we consider steady-state solutions with uniform-curvature, where $\kappa = \kr = \kappa_0$, $\tau$, $\phi^+$ and $\phi^-$ are all constant. Evaluating \rr{nondimeqns} leads to values for the steady-state protein concentrations,
\begin{equation}\label{phi0}
\phi^\pm_0 = \frac{\alpha (1 \pm \tbk0)}{\alpha (1 \pm \tbk0) + 1} 
\end{equation}
and the following transcendental equation for $\kappa_0$:
\begin{equation}\label{transkr}
\kappa_0 - \kappa_c \tanh \left(\frac{2 \alpha \gamma \tbk0}{(\alpha + 1)^2 - \alpha^2 \tbk0^2}\right) =0.
\end{equation}
The straight MT with no differential binding is always a solution to \rr{transkr}, while non-zero roots of $\kappa_0$ correspond to curved states with radius of curvature $\kappa_0^{-1}$. Note that \rr{transkr} depends only on the parameters involved in the protein binding, not those connected to the mechanical response. 

\begin{figure}[thbp]
\begin{overpic}[width=\textwidth,tics=10]{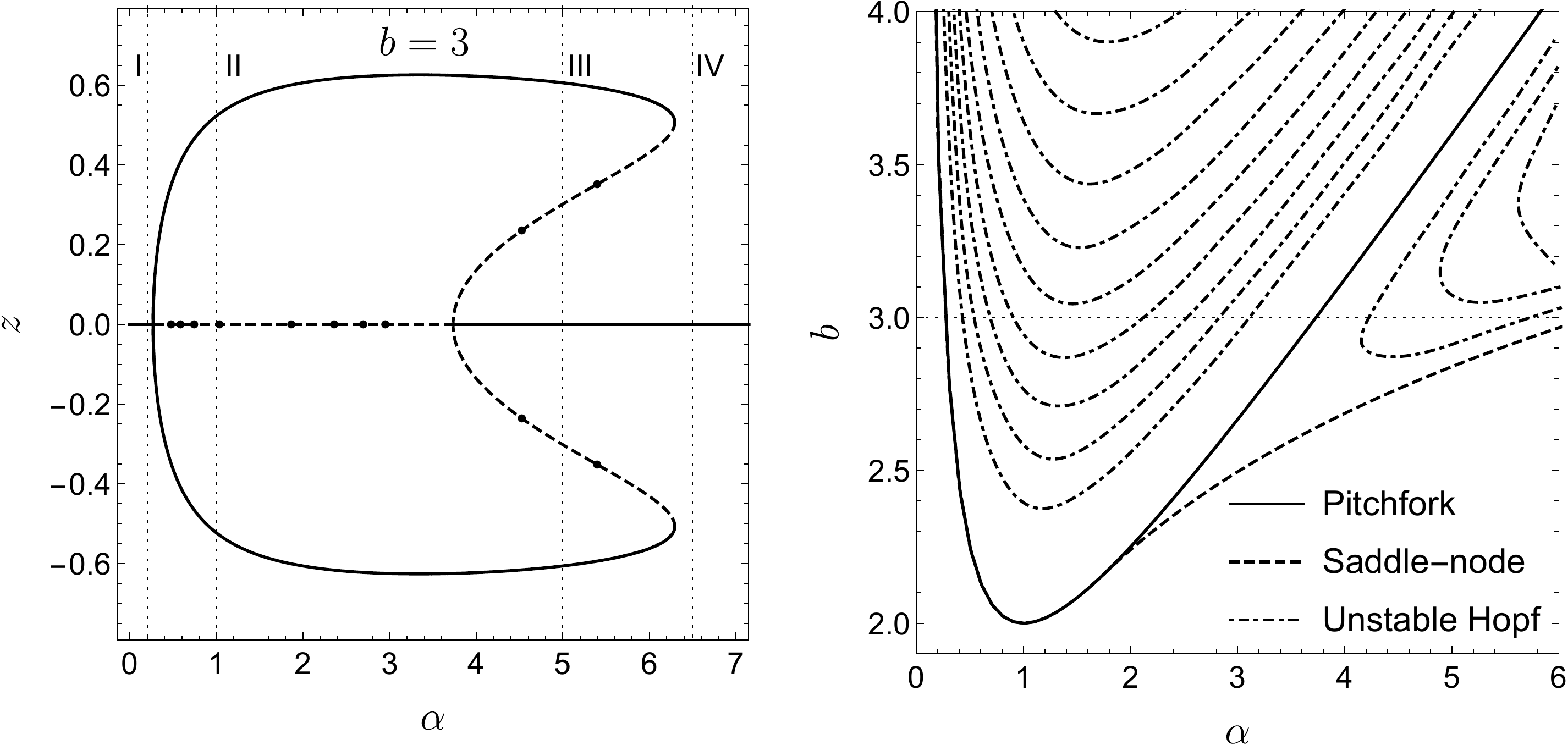}
 \put (0,45) {\Large A}
 \put (50,45) {\Large B}
\end{overpic}
\caption{(A) Bifurcation diagram for $b=3, \gamma=1$, plotting normalised curvature $z$ against dimensionless binding rate $\alpha$. Unstable branches are marked by dashes, dots represent the location of the Hopf bifurcations shown in B. Vertical lines I-IV show positions with 1, 3, 5 and 1 values of $\kappa_0$ respectively. (B) Parameter space map (over binding rate $\alpha$ and preferential binding affinity $b$) showing the position of the three types of local bifurcations seen in the system, for $\gamma=1, \chi_0 = \SI{3}{\micro \metre \tothefourth}, L=\SI{10}{\micro \metre}$. The dotted line corresponds to the bifurcation diagram shown in (A).}
\label{combinedbifhopf}
\end{figure}
Defining $z=\kappa_0 /\kappa_c, b = \beta \kappa_c$, steady-states are associated with the roots of the following three-parameter equation,
\begin{equation}\label{transkrz}
j(z) = z - \tanh \left(\frac{2 \alpha \gamma \tbz}{(\alpha + 1)^2 - \alpha^2 \tbz^2}\right) \equiv z - \tilde{j}(z) =0,
\end{equation}
defining the function $\tilde{j}(z)$. Non-zero roots of $j(z)$ will therefore correspond to steady-states with a non-zero uniform curvature, and so we wish to understand how the parameters affect the existence of these roots. If $j'(0)<0$, then a positive root must exist as $\lim_{z\to\pm \infty} j(z) = \pm \infty$, which leads to a bifurcation condition from where non-zero roots intersect $z=0$,
\begin{equation}\label{bifcon}
j'(0) = \frac{2 \alpha b \gamma}{(\alpha + 1)^2} - 1 = 0.
\end{equation}
As shown in Figure \ref{combinedbifhopf}A, \rr{bifcon} defines two values of the binding rate $\alpha$, ($\alpha_1, \alpha_2$), at which pitchfork bifurcations occur; we focus on $\alpha$ as it is the parameter most available for experimental control.

A linear stability analysis (see Appendix) shows that the uniform solution becomes unstable with a single real positive eigenvalue at $\alpha=\alpha_1$, until it re-stabilizes at $\alpha=\alpha_2$. The first bifurcation at $\alpha=\alpha_1$ is supercritical, producing stable non-uniform solutions, but the solutions arising from $\alpha=\alpha_2$ can be either stable or unstable, depending on the values of $b$ and $\gamma$, with a saddle-node bifurcation occurring further along the branch to re-stabilize solutions when they are initially unstable (Figure \ref{combinedbifhopf}B). This saddle-node bifurcation occurs for a wide range of parameters and leads to multiple non-zero roots as shown in Figure \ref{combinedbifhopf}A. These additional roots appear when $\tilde{j}(z)$ is initially smaller than $z$ but grows sufficiently fast to exceed $z$, giving a second non-zero root. 
 
Provided that $\alpha_1$ and $\alpha_2$ are far enough apart, we also find that a nested series of Hopf bifurcations occur along both the unstable branches as the periodically spaced complex eigenvalues move across the real axis, generating unstable limit cycles (due to the positive eigenvalue still being present), with a corresponding reverse bifurcation when they pass back over the real axis. The positions of these Hopf bifurcations are indicated as points in Figure \ref{combinedbifhopf}A and as curves on Figure \ref{combinedbifhopf}B, which shows the parameter space as $\alpha$ and $b$ are varied for a fixed $\gamma=1$. A similar picture is found as the curvature-binding parameter $\gamma$ is changed, but with a negative relationship between the two feedback parameters $b$ (or $\beta$) and $\gamma$; when one is small the other needs to be large in order for the feedback strength to be large enough to create non-zero roots, as can be seen in \rr{bifcon}.

We have therefore shown there exists a range of parameter values for which multiple stable steady-states exist, allowing for the simultaneous existence of straight and curved MTs at the same parameter values and for a single MT to be transferred between the two when suitably perturbed, as was shown experimentally by \citet{liu2011} (Figure \ref{ringpics}). As can be seen in Figure \ref{combinedbifhopf}A, the non-zero constant value of $\kappa_0$ is generally below the prescribed characteristic curvature $\kappa_c$ (i.e. $z$ is less than 1); this value is approached for some parameter values but can be significantly less, allowing for variation in the ring sizes with a fixed $\kappa_c$. Furthermore, the Hopf bifurcations point to the potential existence of oscillating states, which we will explore below. 

\subsection{Numerical Solutions}
We now solve the full set of partial differential equations, \rr{nondimeqns}-\rr{xydot}, to see how the curvature of the MT evolves in time. To do this, we use the method of lines, discretising in $s$ with a fourth-order central-difference formula to give a set of coupled nonlinear ordinary differential equations in $t$, which are solved using Mathematica. 

\begin{figure}[thbp]
\includegraphics[width=0.95\textwidth,keepaspectratio]{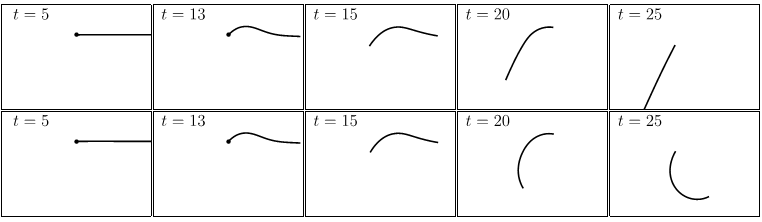} 
\caption{Shape evolution of two MTs which are temporarily pinned before being released. All the parameters are the same except the time of release, which are $t=13, 13.1$ respectively, with $\alpha = 5, b = 3, \gamma = 1, \chi_0 = \SI{3}{\micro \metre \tothefourth}, L=\SI{5}{\micro \metre}$. The resulting end-states correspond to the two stable steady-states seen at line III in Figure \ref{combinedbifhopf}A.}
\label{pinnedshapeevolution}
\end{figure}

As the straight solution is always a steady-state, we need to introduce an initial curvature perturbation to the MT to be able to see other behaviours. One option is to temporarily pin the leading tip of a gliding MT (as is observed in gliding assays when the MT encounters defective motors \citep{bourdieu1995}). For a large enough $\flex$, this will cause the MT to buckle and rotate around the pinned point \citep{sekimoto1995, chelakkot2014,canio2017}. We then allow the MT to unpin after some curvature (and therefore also differential binding) has been generated, and the MT will then either reach a curved configuration or re-straighten. This is shown in Figure \ref{pinnedshapeevolution} and Supplementary Movie S1, where two MTs that are unpinned after slightly different amounts of time settle into the two different steady-states.

Instead of pinning an initially straight MT, we can also generate an initial perturbation by bending the MT into a non-straight configuration and then allowing it to relax, mimicking an interaction with other MTs. In both of these cases, the exact basins of attraction of the two steady-states depends on both the binding and the mechanical parameters; a relatively large perturbation from straight gliding, which persists for long enough to produce binding differences, is required to move the MT into the curved configuration. 

\subsection{Propagating Waves and Chaotic Motion}
\begin{figure}[thbp]
\begin{overpic}[width=0.95\textwidth,tics=10]{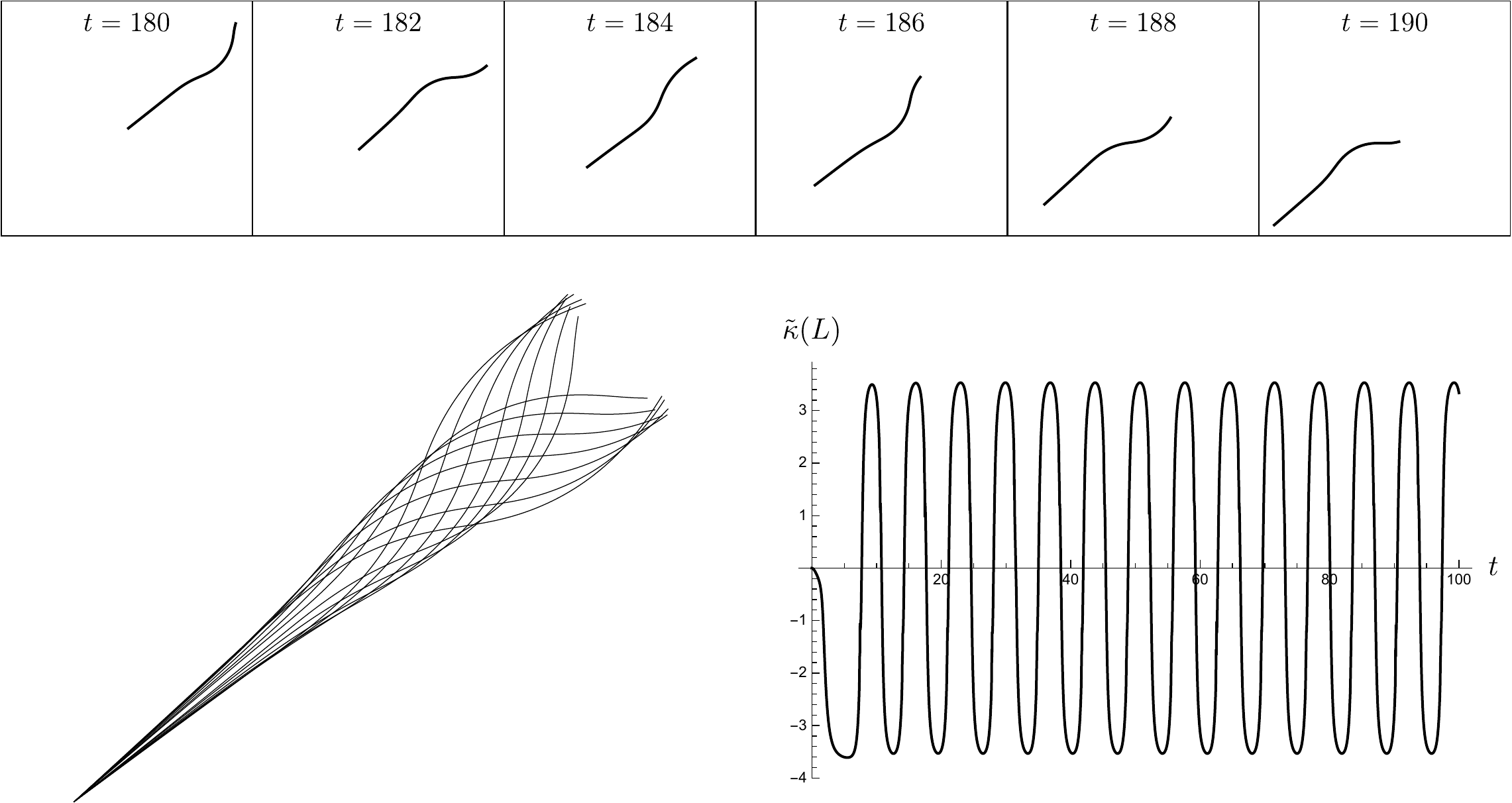}
 \put (-4,52) {\Large A}
 \put (-4,30) {\Large B}
 \put (45,30) {\Large C}
\end{overpic}
\caption{A) Shape evolution of a MT showing oscillatory behaviour as it moves across the surface.  At these parameter values propagating waves in $\kr$ cause the MT to periodically undulate as it moves across the surface, settling into a steady rhythm. One period of these shapes are shown superimposed in (B). For these parameters, the trailing tip preferred curvature, $\kr(L,t)$, plotted in (C), has a regular gap between zeroes of $T=3.466$. Here $\alpha = 8, b=3, \gamma = 1, \chi_0 = \SI{1}{\micro \metre \tothefourth}, L=\SI{5}{\micro \metre}$.}\label{wavyMTs}
\end{figure}
As well as the steady-states detailed in Section \ref{steadystates}, we also find parameter regimes where stable propagating waves move down the MT, shown in Figure \ref{wavyMTs} and Supplementary Movie S2; the resulting MT shapes look strikingly similar to those of cilia beating. These MTs show globally directed motion in the same way as straight MTs, gliding across the surface, but with the addition of a periodic oscillation feeding back from the tip. These sustained oscillations are caused by the reference curvature being translated along the MT as it glides. The curvature of the MT trailing tip, $\kr(L)$, shows the regular periodicity where the waveform reaches the tip (Figure \ref{wavyMTs}B), and we denote the time between zeroes of $\kr(L)$ by $T$. 

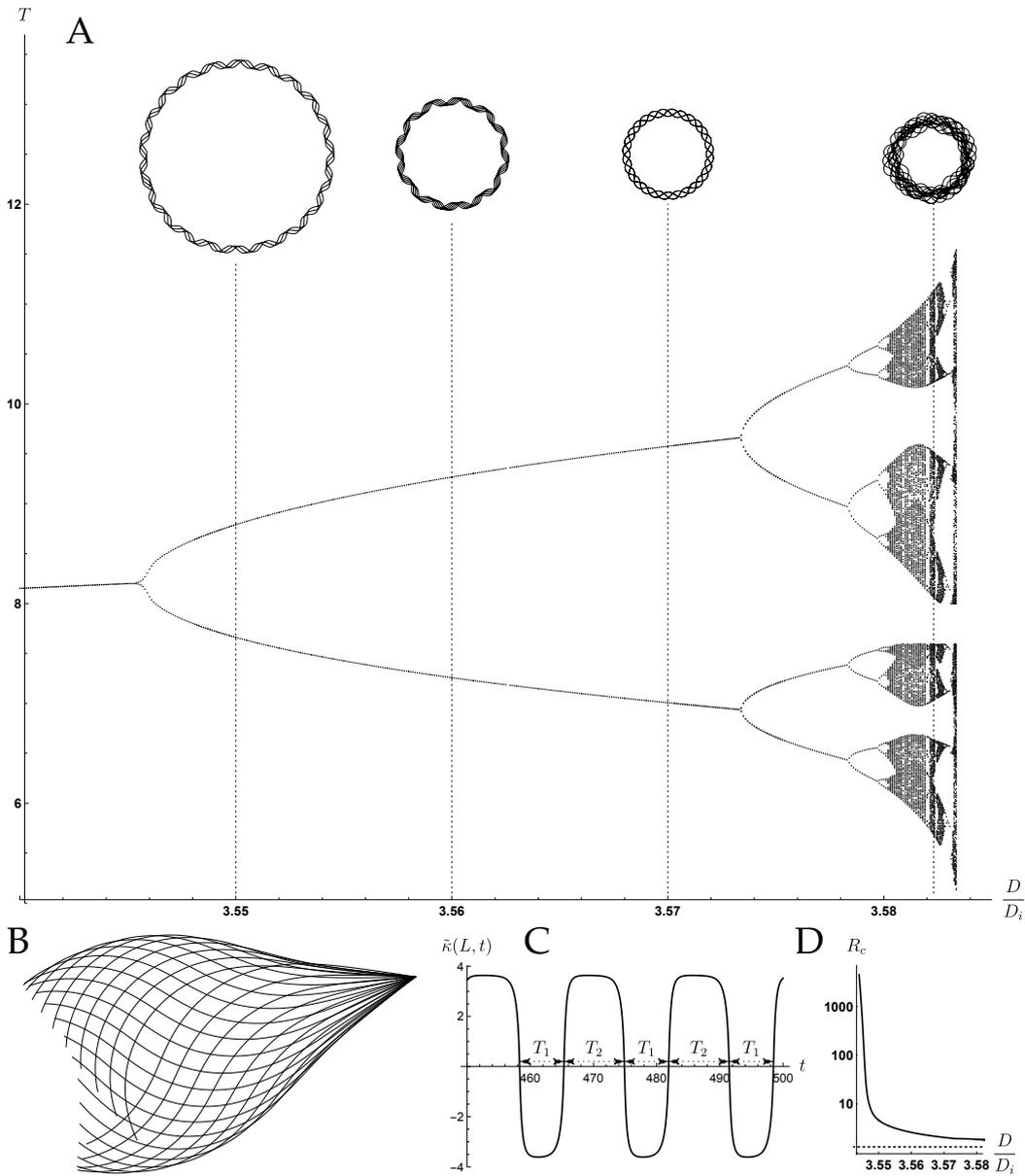
\begin{figure}[thbp]
\begin{overpic}[width=0.95\textwidth,tics=10]{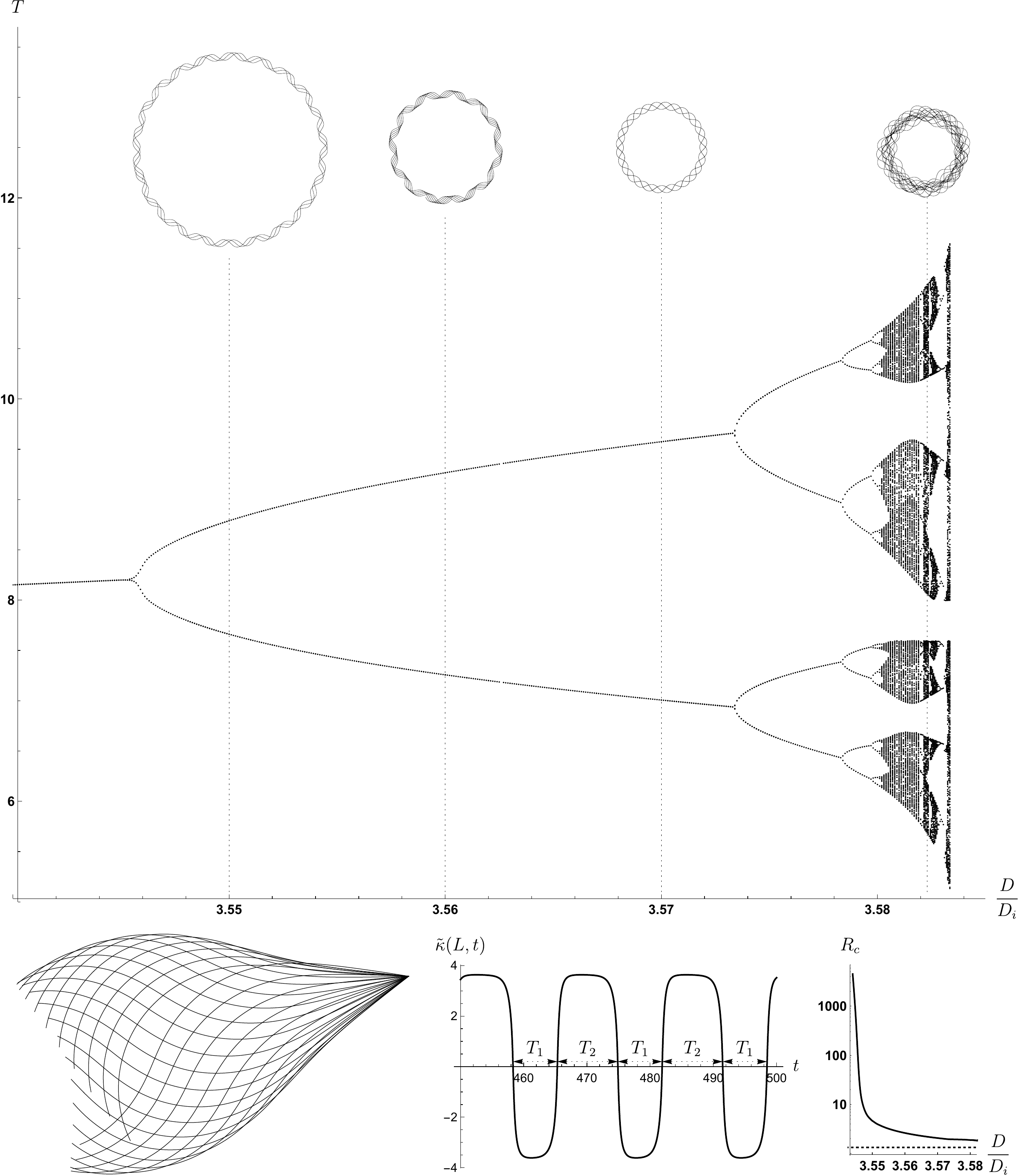}
 \put (5,97) {\Large A}
 \put (0,19) {\Large B}
 \put (44,19) {\Large C}
 \put (67,19) {\Large D}
\end{overpic}
\caption{(A) Time $T$ between the zeroes of the trailing tip curvature $\kr(L,t)$ (after initial transient behaviour) as the diffusion $D$ is increased, showing a period doubling cascade to chaos. Inset above are plots (all on the same scale) showing the motion of the trailing tip of the MT after the initial transient behaviour. For $D/D_i=3.55, 3.56, 3.57$, the MT settles into a steady orbit with decreasing radius, whereas for $D/D_i=3.5823$ the MT moves around the plane in a chaotic manner. (B) Superimposed shapes for $D/D_i=3.57$ over a sub-interval of the orbit, showing a larger amplitude than in Figure \ref{wavyMTs}. (C) Plot of the reference curvature at the end of the MT, $\kr(L)$, showing the two values of $T$ which can be seen in (A). (D) Plot showing how the radius of the large orbit traced out by the oscillating MT decreases monotonically as $D/D_i$ is increased. All other parameters are the same as in Figure \ref{wavyMTs}.}
\label{returnmap}
\end{figure}

As the parameters are changed, the period and amplitude between these oscillations changes smoothly. Additionally, as the diffusion $D$ is increased from $D_i = \SI{0.036}{\micro \metre \squared \per \second}$ \citep{lu2009}, we also see period-doubling bifurcations occur as shown in Figure \ref{returnmap}, where $T$ breaks symmetry, leading to a non-symmetric behaviour in the waveform (Figure \ref{returnmap}C). As $D$ is increased further a period-doubling cascade continues, leading to chaotic behaviour where the MT never settles into a periodic regime, as shown in Figure \ref{returnmap}A. Upon further increasing of $D$, the wave solution disappears and the system settles into the uniform curvature solutions described in Section \ref{steadystates}. While it is unlikely that $D$ could be used as a control parameter in an experiment, these results demonstrate the range of possible outcomes in this system, and their sensitivity to parameter values.

After the initial period-doubling, the overall motion of the MT is biased, leading to the MT moving in a large circle whilst undulating, as shown in the insets of Figure \ref{returnmap}A and in Supplementary Movies S3-S7. The radius of this large circle, $R_c$, decreases monotonically with $D$, as the waveform becomes increasingly asymmetric, shown in Figure \ref{returnmap}D.

These oscillatory solutions, and the period doubling cascade to chaos, appear to exist in regions of parameter space around the second pitchfork bifurcation, $\alpha_2$, shown in Figure \ref{combinedbifhopf}, provided that the \lq wings' of the bifurcation diagram are large enough to include the Hopf bifurcations. They also exist when different sigmoidal functions $\kr = \kappa_c G(\gamma (\phi^+ - \phi^-))$ are used instead of \rr{kreqn}, for instance
\begin{equation}
G(x) = \frac{2}{\pi} \arctan \left(\frac{\pi x}{2}\right), \;\;\; \textrm{erf}\left(\frac{\sqrt{\pi} x}{2}\right), \;\;\; \frac{x}{\sqrt{1+x^2}},
\end{equation}
where $\textrm{erf}$ is the error function (results not shown).

\section{Discussion}
We have presented here a model for the occurrence of rings and arcs seen in MT gliding assays, based on the extrinsic effect of differential binding of the surface-attached kinesins which move the MTs forward. Our model goes beyond that of \citet{ziebert2015}, who assume that the formation of these rings is the result of an intrinsic property of the MTs. Our model incorporates the presence of the kinesins as external factors that drive a feedback loop where MT bending is recognised and stabilised, thus recruiting further kinesins; intrinsic properties of MTs as considered by \citet{ziebert2015} may contribute to these effects and can be considered in our model by modifying the equation for the reference curvature \rr{kreqn}. 

Both models are able to reproduce the key aspect of the experiments, with regimes of bistability where both curved and straight MTs exist simultaneously for the same parameters, differentiated by the local forcing on the MT. In order for rings to be formed, we find that the system needs to be within a parameter regime ($\alpha, \beta, \gamma$) which permits multiple steady-states, and the MT must undergo high induced curvature. Going beyond the model of \citet{ziebert2015}, our approach leads to two predictions which can be experimentally tested:

\begin{enumerate}
\item Our model predicts that varying the effective binding rate $\alpha$ should affect whether or not MT rings can form, as well as their size. This on-rate includes both the binding distance $d$, which can be experimentally altered by truncations of the kinesin-1 tail region as in the experiments performed in \citet{vandelinder2016split}, as well as the surface-bound kinesin density $\Gamma$ which may be directly varied by altering the kinesin concentration. 
\item Our model predicts that high MT density will encourage the formation of rings, consistent with the experiments of \citet{liu2011}. MT density contributes in two ways: it encourages the bending of MTs through MT-MT interactions, and it promotes the pushing down of MTs towards the surface of the assay during cross-over events, significantly enhancing the access to the convex MT side. 
\end{enumerate}

The non-kinesin parameters used in our model are already well-known or easily measurable, but the parameters $\beta$, $\gamma$ and $\kappa_c$ which connect the protein binding to the preferred curvature are entirely unknown. However, they could be characterised via fluid-bending experiments of the kind performed in \citet{peet2017}, and then fitting an appropriate variation of the model described here. The measurements of the curvature sensitivity for the protein doublecortin in taxol-stabilised MTs (Figure 3G of \citet{ettinger2016}) suggest a value of $\beta$ of approximately \SI{3}{\micro \metre}.

Additionally, the exact values for the parameters involved in the combined kinesin on-rate $\alpha$ are not very well characterised. For instance, the surface kinesin density $\Gamma$ is not routinely measured in gliding assays, but can be obtained via landing-rate experiments \citep{katira2007}. Furthermore, other motor proteins than kinesin-1 (e.g. other members of the kinesin superfamily or the minus-end directed dynein/dynactin motor complex), are able to translocate MTs in gliding assays. If parameter changes in these assays facilitate the occurrence of rings, this would provide additional data sets that could be compared and help to refine parameter determination.

The unsteady solutions where the MT oscillates while moving across the surface, including the period-doubling cascade to chaos, are particularly interesting mathematically as they are not immediately obvious from the governing equations. An oscillating regime for a clamped or pinned filament (without the preferred curvature) driven by the tangential motors, as considered here, was found by \citet{sekimoto1995}, with a Hopf bifurcation occurring above a critical forcing $\flex$; \citet{canio2017} also show this behaviour for the case of a filament with a follower force acting on the free end, and expand upon its origin. The behaviour seen here is similar, with a self-sustained oscillatory waveform, but these authors did not report chaotic dynamics. 

These oscillations may be biologically relevant; the wavy MTs generated by our model look similar to those shown in an experimental figure of \citet{gosselin2016}, as well as a MT seen in Supplementary Movie 1 of \citet{scharrel2014}, and similar regularly undulating curved MTs are seen in cells \citep{brangwynne2007, brangwynne2006}; although these are assumed to be caused by mechanical buckling, it is possible that this kind of curvature-dependent binding may enhance the effect. There may also be a connection to the MT phenomenon described as \lq fishtailing', where MTs oscillate laterally whilst their head is stuck, as shown in \citet{applewhite2010} and \citet{weiss1991} for example. We therefore encourage experimentalists to look out for this kind of behaviour.

\section{Future Directions}
Our current model is only a starting point, and a number of further aspects can or should be incorporated.
\begin{enumerate}
\item The arrangement of protofilaments into the MT is via a helical arrangement, with a skew angle inducing a global supertwist for the MTs where moving forward along one protofilament involves rotating around the MT; the exception is for MTs with 13 protofilaments which are almost straight, and this is therefore the type considered in our model so far. MTs with more or fewer protofilaments are shown to rotate as the kinesin moves along them \citep{ray1993}, inducing a torque that could be considered in future versions of the model, particularly as axonal MTs are not always the 13 protofilament type \citep{chaaban2017}. Furthermore, \citet{kawamura2008} find that more rings occur in gliding assays when they use freshly prepared MTs, which have fewer 13-protofilament MTs than their 24 hour aged MTs, suggesting that this MT rotation might enhance the ring-formation by making it easier for the kinesin to be bound to the outer side of the MTs; this effect may be explained by kinesin stepping from one protofilament to another as it reaches its maximum extension, as it does to move around obstacles \citep{schneider2015}.

\item The inclusion of the crosslinking protein streptavidin into gliding assays induces the formation of bundles of curved MTs, known as spools, where multiple MTs are attached together and the entire structure is bent into an arc \citep{vandelinder2016spool}. \citet{luria2011} found more small-circumference spools than predicted by their simulations; these spools have a similar diameter to that of the non-crosslinked MT loops \citep{liu2011, kawamura2008}, and we suggest that the mechanism we propose may be responsible. In particular, \citet{lam2014} found that the size and number of crosslinked spools depends upon the kinesin density, with the presence of more kinesin leading to fewer small-diameter spools, which is consistent with our results. Incorporation of MT-MT interactions via crosslinking proteins would therefore be a natural extension of our model.

\item The core of this model is suitable for other situations where differential protein binding may influence filament curvature, for instance where the protein is freely diffusing in solution as in \citet{peet2017}. The model assumes that the MT stays in the same vertical plane, as it is attached to the surface by the kinesin, and the extension to three dimensions may be required to properly model the situation in cells, incorporating both the twisting as well as the bending of the MT, as well as modelling all the protofilaments individually. 

\item As mentioned in the introduction, other MT-binding proteins occurring in nerve cells, such as tau and doublecortin, have been shown to be curvature-sensitive, and we therefore recommend that these are tested to see if they can also reinforce the curvature. Additionally, an extension to the model to incorporate competition between proteins for the same binding sites may explain how certain proteins have a protective function.

\end{enumerate}

\begin{acknowledgements}
SPP thanks the Leverhulme Trust for the award of an Early Career Fellowship. AP acknowledges the support of the Biotechnology and Biological Sciences Research Council [grant numbers BB/L026724/1, BB/M007553/1, BB/L000717/1].
\end{acknowledgements}

\section*{Appendix: Linear Stability Analysis}\label{linstab}
To investigate the stability of the steady-states, we expand every variable about the constant solution as,
\begin{align*}
\theta &= \Omega t + \kappa_0 s + \epsilon \theta_1(s) e^{\omega t}, \;\; \tau = \epsilon \tau_1(s) e^{\omega t},\\
\phi^{\pm} &= \phi^{\pm}_0 + \epsilon \phi^{\pm}_1(s) e^{\omega t},\;\;
\kr = \kappa_0 + \epsilon \kr_1(s) e^{\omega t},
\end{align*}
where the steady-state values are given by \rr{phi0} and \rr{transkr}, and the angular velocity $\Omega$ is given by
\begin{equation*}\label{rotvel}
\Omega = \xi \chi \flex \kappa_0 = v \kappa_0,
\end{equation*}
connecting the rotation speed of the arc solutions with the free gliding speed.

Linearising with respect to $\epsilon$ and letting $K_1 = \theta_1' -\kr_1$ gives at first order,
\begin{subequations}
\begin{align*}
\omega \theta_1 &= \chi (2 \flex \theta_1' + 3 \kappa_0 \tau_1' + 2 \kappa_0^2 K_1' - K_1''')\\
0 &= \kappa_0^2 \tau_1 - 3 \kappa_0 K_1'' \\
\omega \phi_1^{\pm} &= -(1+ \alpha \pm \alpha \tbk0) \phi^{\pm}_1 \pm \alpha \beta \sech^2(\beta \kappa_0)(1- \phi^{\pm}_0) \kappa_1 - \flex \chi \xi \phi^{\pm \prime}_1 + D \phi^{\pm \prime \prime}_1\\
\kr &= \gamma (\kappa_c + \kappa_0 \tanh(\beta \kappa_0)).
\end{align*} 
\end{subequations}
Writing as a first order matrix system, $\mathbf{y}' = \mathbf{A} \mathbf{y}$ where $\mathbf{y}=(\theta_1,K_1,K_1',K_1'',\tau_1,\tau_1',\phi_1^{+},\phi_1^{+\prime},\phi_1^{-}, \phi_1^{-\prime})$, the matrix $\mathbf{A}$ is given by
\begin{equation*}\nonumber
\mathbf{A} = \left(
\begin{array}{cccccccccc}
0 & 1 & 0 & 0 & 0 & 0 & a_1 & 0 & -a_1 & 0 \\ 
0 & 0 & 1 & 0 & 0 & 0 & 0 & 0 & 0 & 0 \\ 
0 & 0 & 0 & 1 & 0 & 0 & 0 & 0 & 0 & 0 \\ 
 - \omega \chi^{-1} & \xi \flex & \xi \kappa_0^2 & 0 & 0 & (1+\xi)\kappa_0 & \xi \flex a_1 & 0 & -\xi \flex a_1 & 0 \\ 
0 & 0 & 0 & 0 & 0 & 1 & 0 & 0 & 0 & 0 \\ 
0 & 0 & 0 & -(1+\xi)\kappa_0 \xi^{-1} & \kappa_0^2 \xi^{-1} & 0 & 0 & 0 & 0 & 0 \\ 
0 & 0 & 0 & 0 & 0 & 0 & 0 & 1 & 0 & 0 \\ 
0 & -a_2^{+} & 0 & 0 & 0 & 0 & a_3^{+} + \omega D^{-1}& -\flex \xi \chi D^{-1}  & \gamma a_1 a_2^{+} & 0 \\ 
0 & 0 & 0 & 0 & 0 & 0 & 0 & 0 & 0 & 1 \\ 
0 & a_2^{-} & 0 & 0 & 0 & 0 & a_1 a_2^{-} & 0 & a_3^{-} +\omega D^{-1}& -\flex \xi \chi D^{-1} \\ 
\end{array}
\right) 
\end{equation*}
where 
\begin{align*}
a_1 &= \gamma \left(\kappa_c + \kappa_0 \tanh\left( \frac{2\alpha \gamma \tbk0}{\alpha^2 \tbk0^2-(1+\alpha)^2}\right) \right)\\
a_2^{\pm} &= \frac{\alpha \beta \sech(\beta \kappa_0)^2}{D(1+\alpha \pm \alpha \tbk0)}\\
a_3^{\pm} &= \frac{1+ \alpha (2-a_1 \beta \gamma) + \alpha^2 \pm 2 \alpha (1+\alpha)\tbk0 + \alpha (\alpha + a_1\beta \gamma)\tbk0^2}{D (1+ \alpha \pm \alpha \tbk0)}.
\end{align*}
We find the eigenvalues of this linear boundary-value problem using the Compound Matrix method to calculate the Evans function \citep{afendikov2001} in Mathematica (a package implementation of which is available from github.com/SPPearce/CompoundMatrixMethod) and chebfun \citep{driscoll2014} in Matlab. The spectrum of eigenvalues here is discrete, with regularly spaced pairs of complex conjugates.

\section*{Appendix: Supplementary Video Captions}

Supplementary Video 1: Shape evolution of two MTs which are temporarily pinned before being released. All the parameters are the same except the time of release, which are $t=13, 13.1$ respectively, with $\alpha = 5, b = 3, \gamma = 1, \chi_0 = \SI{3}{\micro \metre \tothefourth}, L=\SI{5}{\micro \metre}$. Corresponds to the still images in Figure 4. 
\\
Supplementary Video 2: Shape evolution of a MT showing oscillatory behaviour as it moves across the surface. Here $\alpha = 8, b=3, \gamma = 1, \chi_0 = \SI{1}{\micro \metre \tothefourth}, L=\SI{5}{\micro \metre}$, and corresponds to the extracted frames in Figure 5.
\\
Supplementary Video 3: Shape evolution of a MT showing oscillatory behaviour, before the period doubling bifurcation, with $D=3.5 D_i$. Here $\alpha = 8, b=3, \gamma = 1, \chi_0 = \SI{1}{\micro \metre \tothefourth}, L=\SI{5}{\micro \metre}$. 
\\
Supplementary Video 4: Shape evolution of a MT showing oscillatory behaviour, past the period doubling bifurcation, with $D=3.55 D_i$. Here $\alpha = 8, b=3, \gamma = 1, \chi_0 = \SI{1}{\micro \metre \tothefourth}, L=\SI{5}{\micro \metre}$. Corresponds to the first vertical dotted line in Figure 6. 
\\
Supplementary Video 5: Shape evolution of a MT showing oscillatory behaviour, past the period doubling bifurcation, with $D=3.56 D_i$. Here $\alpha = 8, b=3, \gamma = 1, \chi_0 = \SI{1}{\micro \metre \tothefourth}, L=\SI{5}{\micro \metre}$. Corresponds to the second vertical dotted line in Figure 6.
\\
Supplementary Video 6: Shape evolution of a MT showing oscillatory behaviour, past the period doubling bifurcation, with $D=3.57 D_i$. Here $\alpha = 8, b=3, \gamma = 1, \chi_0 = \SI{1}{\micro \metre \tothefourth}, L=\SI{5}{\micro \metre}$. Corresponds to the third vertical dotted line in Figure 6.
\\
Supplementary Video 7: Shape evolution of a MT showing oscillatory behaviour, past the period doubling bifurcation, with $D=3.5823 D_i$. Here $\alpha = 8, b=3, \gamma = 1, \chi_0 = \SI{1}{\micro \metre \tothefourth}, L=\SI{5}{\micro \metre}$. Corresponds to the fourth vertical dotted line in Figure 6.

\bibliographystyle{chicago}
\bibliography{MicrotubuleReferences}
\end{document}